\def\be{\begin{equation}}
\def\ee{\end{equation}}
\def\bea{\begin{eqnarray}}
\def\eea{\end{eqnarray}}
\def\bse{\begin{subequations}}
\def\ese{\end{subequations}}

\documentclass[prl,twocolumn,showpacs,preprintnumbers,amsmath,amssymb]{revtex4}
\usepackage{graphicx}
\usepackage{dcolumn}
\usepackage{bm}
\begin{document}
\preprint{NSF-KITP-05-31}
\title{Blue Quantum Fog: Chiral Condensation in Quantum Helimagnets\\
\vskip 1mm
}
\author{Sumanta Tewari$^{1,2}$, D. Belitz$^{1,3}$, and
        T.R. Kirkpatrick$^{1,2}$}
\affiliation{$^{1}$Kavli Institute for Theoretical Physics,
University of
California, Santa Barbara, CA 93106\\
             $^{2}$Institute for Physical Science and Technology and Department of
                   Physics, University of Maryland, College Park, MD 20742\\
         $^{3}$Department of Physics and Materials Science Institute, University
                of Oregon, Eugene, OR 97403}
\date{\today
}
\begin{abstract}
It is shown that a condensation transition involving a chiral order parameter
can occur in itinerant helimagnets, in analogy to the transition between the
isotropic phase and the phase known as blue fog or blue phase III in
cholesteric liquid crystals. It is proposed that such a transition is the
explanation for recent neutron scattering results in MnSi. Predictions are made
that will allow to experimentally test this proposal.
%
\end{abstract}

\pacs{}

\maketitle

The unusual behavior of the low-temperature itinerant magnet MnSi has received
much attention. At ambient pressure $P$, the material enters a magnetic phase
below a temperature $T_c \approx 30{\text K}$. Over distances of a few lattice
spacings, the magnetic order appears ferromagnetic
\cite{Pfleiderer_et_al_1997}. However, at longer length scales a helical
modulation of the magnetization appears, with a wavelength $2\pi/Q_0 \approx
170{\text\AA}$ \cite{Ishikawa_et_al_1976}. This helical structure is believed
\cite{Bak_Jensen_1980, Nakanishi_et_al_1980} to be due to an interaction
between magnetic fluctuations ${\bm M}$ of a form first proposed by
Dzyaloshinski \cite{Dzyaloshinski_1958} and Moriya \cite{Moriya_1960} (DM),
$\int d{\bm x}\ {\bm M}\cdot({\bm \nabla}\times{\bm M})$. Such a term, which is
produced by the spin-orbit interaction, is allowed by symmetry in MnSi since
its lattice structure lacks inversion symmetry. The helix is readily observed
via neutron scattering, with scattering intensity appearing only in the
$\langle 111\rangle$ direction since crystal fields lock the direction of the
helix \cite{Ishikawa_et_al_1976, Pfleiderer_et_al_2004}. With increasing $P$,
the transition temperature to the ordered phase decreases monotonically, and
the nature of the transition changes from continuous or very weakly first order
to decidedly first order at a tricritical point at $P^*\approx
12\,{\text{kbar}}$, before the transition temperature drops to zero at
$P_c\approx 14.6\,{\text{kbar}}$. In the disordered phase, i.e., for
$P\!>\!P_c$, spectacular non-Fermi-liquid (NFL) behavior of the transport
properties has been observed below a crossover temperature $T^*$
\cite{Pfleiderer_Julian_Lonzarich_2001, Doiron-Leyraud_et_al_2003}.

Within this extended NFL region, Pfleiderer {\it et al.}
\cite{Pfleiderer_et_al_2004} have recently identified a pressure-dependent
temperature scale, $T_0$, below which there is a strong neutron scattering
signal at a well-defined wave number $q_0 \approx 0.043\,{\text{\AA}}$. The
signal is strong enough to be reminiscent of the one at $Q_0 \approx
0.037{\text\AA}$ in the ordered phase, but is much more isotropic, with broad
maxima centered around the $\langle 110\rangle$ direction. Reference
\cite{Pfleiderer_et_al_2004} has interpreted these observations as evidence for
the existence of short-ranged helical order even in the non-magnetic phase, and
has suggested that this short-ranged helical order is at the heart of the NFL
transport behavior.

\begin{figure}[b]
\vskip -0mm
\includegraphics[width=7.0cm]{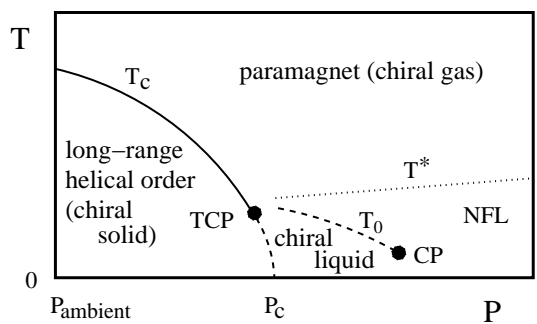}
\caption{Phase diagram of MnSi, including the proposed chiral liquid phase with
an associated critical point (CP). The CP may be at $T<0$ and thus be
inaccessible. Second (or very weakly first) order transitions, first-order
transitions, and crossovers are denoted by solid, dashed, and dotted lines,
respectively. The non-Fermi-liquid (NFL) region includes the chiral liquid, and
the chiral gas extends below $T^*$. The precise structure near the tricritical
point (TCP) is not known. See the text for further information.}
\label{fig:1}
\end{figure}

In this Letter we focus on such local correlations and argue that the existence
of the temperature scale $T_0$ is consistent with a first-order transition from
a chiral liquid to a gaseous phase as one crosses the condensation temperature
$T_0(P)$. We thus propose that the phase diagram of MnSi is more complicated
than previously thought, with a liquid-gas-type transition inside the
non-magnetic phase. This proposed phase diagram is schematically depicted in
Fig.\ \ref{fig:1}.
A convenient order parameter for this first-order transition is the chiral
composite field $\psi = {\bm M}\cdot({\bm \nabla}\times{\bm M})$. $\psi$ is a
pseudoscalar which has nonvanishing average values both below and above $T_0$.
The two phases separated by $T_0$ thus have the same symmetry, as do the
gaseous and liquid phases, respectively, of ordinary fluids. Crossing the
coexistence line $T_0(P)$ is accompanied by a discontinuous change in the
expectation value of $\psi$, which corresponds to stronger short-ranged helical
correlations in the liquid than in the gas. This accounts for a stronger
neutron scattering signal in the liquid, with the signal being isotropic to
zeroth approximation. Our scenario is analogous to the transition from the
isotropic phase to the phase known as `blue phase III' or `blue fog' in
cholesteric liquid crystals, for which a theory based on liquid-gas type first
order transition is the currently most successful interpretation
\cite{Lubensky_Stark_1996, Anisimov_Agayan_Collings_1998, Englert_et_al_2000,
Longa_Ciesla_Trebin_2003, Ciesla_Longa_2004}. This condensation interpretation
makes `blue fog' a very appropriate name for the phase below the coexistence
temperature. We will first show, starting from a microscopic quantum mechanical
action for a helimagnet, that there is an attractive interaction between chiral
fluctuations described by the order parameter $\psi$, which makes a
gas-liquid-like condensation of these degrees of freedom possible for
appropriate values of $T$ and $P$. Assuming that this condensation can be
identified with the observed temperature scale $T_0(P)$, we will focus on
experimental predictions that follow from this proposal.

Let us start with a Landau-Ginzburg-Wilson (LGW) functional appropriate for a
quantum helimagnet that adds a DM interaction to a ferromagnetic action:
\bse
\label{eqs:1}
\be
\hskip -0pt S_{\text{DM}}[{\bm M}] = S_{\text{fm}}[{\bm M}] + c\int dx\, {\bm
M}(x)\cdot\left[{\bm\nabla}\times{\bm M}(x)\right],
\label{eq:1a}
\ee
\vskip -5mm
\bea
S_{\text{fm}}[{\bm M}] &=& \frac{1}{2} \int dx\,dy\ {\bm
M}(x)\,\Gamma(x-y)\,{\bm M}(y)
\nonumber\\
 && +\, \frac{u}{4}\int dx\, \left({\bm M}^2(x)\right)^2.
\label{eq:1b}
\eea
The two-point vertex $\Gamma$ reads, in Fourier space,
\be
\Gamma({\bm p},i\omega_n) = t + a\,{\bm p}^2 + b\,\vert\omega_n\vert/\vert{\bm
p}\,\vert.
\label{eq:1c}
\ee
\ese
Here $x = ({\bm x},\tau)$ is a four-vector that comprises position ${\bm x}$
and imaginary time $\tau$, and $\int dx = \int_V d{\bm x}\int_0^{1/T}d\tau$.
${\bm p}$ is a wave vector, and $\omega_n = 2\pi Tn$ denotes a bosonic
Matsubara frequency. ${\bm M}$ is the order parameter field with components
$M_i$ ($i=1,2,3$) whose expectation value is proportional to the magnetization.
$t$, $a$, $b$, $c$, and $u$ are the parameters of the LGW theory; they are
functions of $T$ and $P$. $V$ is the system volume. $S_{\text{fm}}$ is Hertz's
action for a quantum ferromagnet \cite{Hertz_1976, nonanalyticity_footnote}.
The additional term in $S_{\text{DM}}$ is the chiral DM term.
The sign of $c$ determines the handedness of the helix; we will take $c>0$
without loss of generality.

For the Gaussian propagator $G_{ij}({\bm p},i\omega_n)\equiv \langle M_i({\bm
p},i\omega_n) M_j(-{\bm p},-i\omega_n)\rangle$ we obtain from Eqs.\
(\ref{eqs:1})
\bea
G_{ij}({\bm p},i\omega_n) &=&  \frac{1}{\Gamma^2({\bm p},i\omega_n) - c^2{\bm
p}^{\,2}}\,\Bigl[\Gamma({\bm p},i\omega_n)\,\delta_{ij}
\nonumber\\
&& \hskip 15pt - i c\,p_{\,l}\,\epsilon_{ij\,l} - c^2 p_i p_j/\Gamma({\bm
p},i\omega_n) \Bigr].
%
\label{eq:2}
\eea
An analysis of the eigenvalue problem given by the quadratic form $G_{ij}$
shows that the paramagnetic phase is unstable against the formation of helical
order for $t < c^2/4a$. The instability first occurs at $\vert{\bm p}\,\vert =
q_0 = c/2a$, the pitch of the helix. $t(T_c,P) = c^2/4a$ thus determines $T_c$
in Fig.\ \ref{fig:1} in a mean-field approximation.

We now consider the completeness of the action given by Eqs.\ (\ref{eqs:1}).
The only chiral term so far is the quadratic-in-${\bm M}$ DM term with coupling
constant $c$, and the action $S_{\text{DM}}$ is the magnetic analog of the
action for cholesteric liquid crystals that was the starting point for the
theory developed by Lubensky and Stark \cite{Lubensky_Stark_1996}. However,
symmetry allows for a quartic chiral term of the form $\int dx\ \left({\bm
M}\cdot({\bm \nabla}\times{\bm M})\right)^2$. It is easy to see that such a
term is indeed generated by a perturbative renormalization-group procedure,
with a {\em negative definite} coupling constant $0
> -d_1 \propto -u^2\,c^2$. Another quartic term that is allowed by symmetry and
generated by renormalizing the action $S$ is $-d_2\int dx\ {\bm M}^2\left({\bm
M}\cdot({\bm\nabla}\times{\bm M})\right)$, with $0
> -d_2 \propto - u^2c$. Adding these two terms to Eq.\ (\ref{eq:1a}), we obtain
our final LGW action, with $d_1, d_2 > 0$,
\bea
S[{\bm M}] &=& S_{\text{DM}}[{\bm M}] - d_1\int dx\,\left[{\bm
M}(x)\cdot\left({\bm\nabla}\times{\bm M}(x)\right)\right]^2
\nonumber\\
&& - d_2\int dx\, {\bm M}^2(x)\,\left[{\bm
M}(x)\cdot\left({\bm\nabla}\times{\bm M}(x)\right)\right].
\label{eq:3}
\eea

The $d_1$-term is conceptually crucial for the physical picture we are
proposing. The DM term ensures a nonzero expectation value $\langle{\bm
M}\cdot({\bm\nabla}\times{\bm M})\rangle\neq 0$ everywhere in the phase
diagram.
The presence of $d_1 > 0$ implies an {\em attractive} interaction between the
chiral fluctuations described by ${\bm M}\cdot({\bm\nabla}\times{\bm M})$. This
in turn means that the chiral fluctuations may condense into a chiral liquid as
the temperature is lowered, with a discontinuous behavior of $\langle {\bm
M}\cdot({\bm\nabla}\times{\bm M})\rangle$ across a first-order phase
transition.
This is the central idea of the present paper.

The above considerations suggest considering the composite field $\psi = {\bm
M}\cdot({\bm\nabla}\times{\bm M})$ an order parameter for a possible chiral
first-order phase transition. It is thus desirable to construct an effective
action in terms of $\psi$. The simplest way to do this is to perform a
Hubbard-Stratonovich decoupling of the combined $d_1$ and $d_2$-terms in the
action, Eq.\ (\ref{eq:3}), with $\psi(x)$ the auxiliary field. Alternatively,
one can constrain ${\bm M}\cdot({\bm\nabla}\times{\bm M})$ to $\psi$ by means
of a Lagrange multiplier field that is later integrated out
\cite{Lubensky_Stark_1996}. Either method yields an effective action which
contains all terms allowed by symmetry, and which leads to the same partition
function as the original action. If one integrates out ${\bm M}$, one obtains
an action in terms of $\psi$ alone, which is of the form of an LGW action that
describes a liquid-gas transition \cite{Chaikin_Lubensky_1995}. This is not
very illuminating, since
the $\psi$-correlation functions are not directly measurable. It is therefore
advantageous to integrate out only the `fast' (i.e., large-momentum and
high-frequency) components of the field ${\bm M}$ and write the theory in terms
of $\psi$ and the slow components of ${\bm M}$, whose correlation functions
{\em are} directly measurable. If we denote the slow components of ${\bm M}$ by
the same symbol for simplicity, we thus obtain the following final effective
action for chiral fluctuations $\psi$ and slow magnetization fluctuations ${\bm
M}$,
\be
S_{\text{eff}}[{\bm M},\psi]= S_{\text{DM}}[{\bm M}] + S_{\psi}[\psi] +
S_{\text c}[{\bm M},\psi].
\label{eq:4}
\ee
The first part of the effective action has the same functional form as the
action $S_{\text{DM}}$ given by Eqs.\ (\ref{eqs:1}), only the parameters have
different values. The chiral part, $S_{\psi}[\psi]$, is an LGW functional for a
scalar order parameter with no invariance under the transformation $\psi\to
-\psi$,
\bea
S_{\psi}[\psi] &=& \int dx\,\Bigl[ r\,\psi^2(x)
 - h\,\psi(x) + s\,\vert{\bm \nabla} \psi(x)\vert^2
 - v\,\psi^3(x)
\nonumber\\
&& \hskip 50pt +\ w\, \psi^4(x)\Bigr].
\label{eq:5}
\eea
For suitable parameter values, $\psi$ can thus undergo a first-order phase
transition. Finally, the coupling term reads
\bea
S_{\text c}[{\bf M},\psi] &=& \int dx\,\Bigl[ g_1\,{\bf M}(x)\cdot\left[{\bm
\nabla}\times {\bf M}(x)\right]\psi(x)
\nonumber\\
&& \hskip 20pt + g_2\,{\bm M}^2(x)\,\psi(x)\Bigr].
\label{eq:6}
\eea
$g_1>0$, $g_2>0$ have the same sign as $d_1$, $d_2$. The coupling constants
$r$, $h$, $s$, etc.,
are functions of temperature and pressure, as are the coupling constants of the
starting LGW theory in Eqs.\ (\ref{eqs:1}). The structures of all terms in the
action $S_{\text{eff}}$ are governed by symmetry requirements. Once one has
introduced the pseudoscalar order parameter $\psi$ in addition to ${\bm M}$,
one therefore can in principle just write down $S_{\text{eff}}$ based on
symmetry considerations.

In what follows, we will assume that the first-order transition inherent in the
theory occurs in the experimentally accessible range of $P$ and $T$ and can be
identified with the observed temperature scale $T_0$. We will now discuss some
simple observable consequences of this proposal. For simplicity, we will treat
$\psi$ in mean-field approximation, $\psi(x)\approx \langle\psi(x)\rangle
\equiv\psi = {\text{const.}}$ $\psi$ increases discontinuously as one crosses
the coexistence curve $T_0(P)$ from above, and the discontinuity goes to zero
as one approaches the critical point that marks the end of the coexistence
curve $T_0(P)$, see Fig.\ \ref{fig:1}.

Observable consequences of the first-order transition arise from the coupling
of the chiral order parameter $\psi$ to the magnetization via Eq.\
(\ref{eq:6}). In our mean-field approximation, which treats $\psi$ as a
constant, this coupling simply renormalizes the terms quadratic in ${\bm M}$ in
the action $S_{\text{DM}}$, leading to renormalized coupling constants
\bse
\label{eqs:7}
\be
c_{\text R} = c + g_1\psi \quad,\quad t_{\text R} = t + g_2\psi,
\label{eq:7a}
\ee
and a renormalization of the vertex $\Gamma$, Eq.\ (\ref{eq:1c}), given by
\be
\Gamma_{\text R}({\bm p},i\omega_n) = t_{\text R} + a\,{\bm p}^{\,2} +
b\,\vert\omega_n\vert/\vert{\bm p}\,\vert.
\label{eq:7b}
\ee
Since the proposed first-order transition occurs within the magnetically
disordered phase, the terms of higher order in ${\bm M}$ are not qualitatively
important; they can be treated perturbatively and lead to further
renormalizations of the Gaussian action. The physical magnetic susceptibility
tensor $\chi_{ij}({\bm p},i\omega_n) = \langle M_i({\bm p},\omega_n)\,M_j(-{\bm
p},-\omega_n)\rangle$ therefore has the same form as the Gaussian propagator
given by Eq.\ (\ref{eq:2}), but it now depends on the chiral order parameter
$\psi$,
\bea
\chi_{ij}({\bm p},i\omega_n) &=& \frac{1}{\Gamma_{\text R}^2({\bm p},i\omega_n)
- c_{\text R}^2{\bm p}^{\,2}}\,\Bigl[
   \Gamma_{\text R}({\bm p},i\omega_n)\,\delta_{ij}
\nonumber\\
&&\hskip -10pt - i c_{\text R}\,p_{\,l}\,\epsilon_{ij\,l}
   - c_{\text R}^2 p_i p_j/\Gamma_{\text R}({\bm p},i\omega_n)\Bigr].
\label{eq:7c}
\eea
\ese
In particular, the thermodynamic magnetic susceptibility $\chi_{ij} =
\delta_{ij}\,\chi_{\text m}$, defined as $\chi_{ij} = \lim_{{\bm p}\to 0}\int
d\omega\,{\text{Im}} \chi_{ij}({\bm p},\omega + i0)/\omega$ is
$\psi$-dependent,
\be
\chi_{\text m} = 1/t_{\text R}.
\label{eq:8}
\ee

We now discuss these results. The energy-resolved neutron scattering
cross-section $d^{\,2}\sigma/d\Omega\,d\omega$, with $\Omega$ the solid angle
and $\omega$ the frequency or energy, is related to the quantity $\chi({\bm q})
\equiv (\delta_{ij} - {\hat q}_i {\hat q}_j)\chi_{ij}({\bm q},i\omega_n=0)$ by
\cite{Chaikin_Lubensky_1995}
\be
\chi({\bm q}) = {\text{const.}}\times\int_{-\infty}^{\infty}
d\omega\,\frac{1}{\omega}\,\left(1 -
e^{-\omega/T}\right)\,\frac{d^{\,2}\sigma}{d\Omega\, d\omega}.
\label{eq:9}
\ee
From Eq.\ (\ref{eq:7c}) one finds for this weighted frequency average of the
cross-section
\be
\chi({\bm q}) = 2\,\frac{t_{\text R} + a\,{\bm q}^2}{\left(t_{\text R} +
a\,{\bm q}^2\right)^2 - c_{\text R}^2{\bm q}^2}.
\label{eq:10}
\ee
For $c_{\text R}^2/4a < t_{\text R} < c_{\text R}^2/a$ the system is in the
disordered phase and $\chi({\bm q})$ has a maximum at $q = q_0 > 0$ with
\be
a\,q_0^2 = c_{\text R}\sqrt{t_{\text R}/a} - t_{\text R}.
\label{eq:11}
\ee
A measure for the height of the peak at $q_0$ is
\be
\chi(q_0)/\chi(0) = 1/(2\sqrt{y} - y),
\label{eq:12}
\ee
which depends only on $y = c_{\text R}^2/t_{\text R}a$. The peak is higher and
sharper for smaller $t_{\text R}$ (at fixed $c_{\text R}$).

The theory thus yields, in the disordered phase not too far from the boundary
to long-range magnetic order, a sharp peak in $\chi({\bm q})$ at a wave number
on the same order as the pitch of the helix in the ordered phase, in
qualitative agreement with experiment \cite{Pfleiderer_et_al_2004,
neutron_scattering_footnote}. Upon crossing the first-order transition from
chiral liquid to the gas, $\psi$ decreases and the peak in $\chi({\bm q})$
becomes less pronounced \cite{peak_footnote}, again in qualitative agreement
with the interpretation of $T_0$ as a chiral first-order transition.

Two predictions of other observable effects are: (1) A latent heat $Q$ across
the coexistence line $T_0(P)$.
The absolute value of $Q$ will be small, since the transition takes place at
low temperatures. (2) A discontinuity of the magnetic susceptibility
$\chi_{\text m}$ across $T_0(P)$, see Eq.\ (\ref{eq:8}), even though no
magnetic transition occurs at that temperature.
Again, this effect will be small since $g_2$ must be small in order for the
first-order transition to occur in the first place \cite{peak_footnote}. If
such signs of a first-order transition are observed, it would also be
worthwhile to look for the critical point (CP in Fig.\ \ref{fig:1}), which will
be characterized by critical fluctuations in the Ising universality class.

The present theory
cannot easily explain the magnitude of the increase in the neutron scattering
observed below $T_0$. It is worth noting that the original form of the
Lubensky-Stark theory for liquid crystals \cite{Lubensky_Stark_1996} also did
not explain the huge increase observed by light scattering in the blue fog
phase. Later theories that took into account more sophisticated correlation
effects did, however, find anomalously large fluctuations
\cite{Englert_et_al_2000, Longa_Ciesla_Trebin_2003, Ciesla_Longa_2004}, and we
expect the same to be true for the present theory. A quantitative understanding
of the anomalously large scattering below $T_0$ may also be needed to
understand the NFL transport behavior mentioned in the introduction, and the
size of the liquid region in the phase diagram.

Finally, we need to discuss the relation between the present theory and related
treatments of chiral liquid crystals, as the very possibility of such a
relation has been questioned. Early work on blue phases in liquid crystals
focused on double-twist cylinder configurations of the director. In this
context Wright and Mermin \cite{Wright_Mermin_1989} have argued that there
cannot be analogs of blue phases in helical magnets, and they bolstered this
argument by free-energy considerations at the mean-field level. These arguments
do not apply to our proposal, for several reasons. (1)
It has since become clear that blue phase III
needs to be considered separately from the other blue phases. Double-twist
cylinder configurations are not central to the current understanding of blue
phase III; they are just one of many speculations concerning local order in
this phase \cite{Crooker_2001}. (2) The free-energy argument of Ref.\
\onlinecite{Wright_Mermin_1989} is most applicable to the crystalline blue
phases I and II
because entropic contributions that arise in disordered phases are not taken
into account. (3) The free-energy argument breaks down when the terms with
coupling constants $d_1$, $d_2$
are included in the
free-energy functional.

Given that a magnetic analog of the blue fog phase or blue phase III cannot be
ruled out, an obvious first step is to construct a theory analogous to the one
put forward in Ref.\ \onlinecite{Lubensky_Stark_1996}, which is what we have
done above. There are two main differences between the two theories. One is the
purely technical point that we deal with a quantum vector order parameter (the
magnetization) instead of a tensor classical one (the director). The other is
the existence in our theory of the couplings $d_1$ and $d_2$ in Eq.\
(\ref{eq:3}). These terms are allowed by symmetry and also arise in the
explicit derivation sketched above. The attractive sign of $d_1$ is crucial,
since it forms the physical basis for a condensation scenario. In the final
effective action, these terms give rise to the coupling constants $g_1$ and
$g_2$ in Eq.\ (\ref{eq:6}). The analog of the former was also present in Ref.\
\onlinecite{Lubensky_Stark_1996}, although its sign was not obvious.

This work was initiated at the Aspen Center for Physics. We thank the
participants of the Workshop on Quantum Phase Transitions at the KITP at UCSB,
and in particular Christian Pfleiderer, for stimulating discussions. S.T. would
like to thank the Institute for Theoretical Physics at the University of Oregon
for hospitality. This work was supported by the NSF under grant Nos.
DMR-01-32555, DMR-01-32726, and PHY99-07949.

\vskip -6mm

\begin{thebibliography}{24}
\expandafter\ifx\csname natexlab\endcsname\relax\def\natexlab#1{#1}\fi
\expandafter\ifx\csname bibnamefont\endcsname\relax
  \def\bibnamefont#1{#1}\fi
\expandafter\ifx\csname bibfnamefont\endcsname\relax
  \def\bibfnamefont#1{#1}\fi
\expandafter\ifx\csname citenamefont\endcsname\relax
  \def\citenamefont#1{#1}\fi
\expandafter\ifx\csname url\endcsname\relax
  \def\url#1{\texttt{#1}}\fi
\expandafter\ifx\csname urlprefix\endcsname\relax\def\urlprefix{URL }\fi
\providecommand{\bibinfo}[2]{#2} \providecommand{\eprint}[2][]{\url{#2}}

\bibitem[{\citenamefont{Pfleiderer et~al.}(1997)\citenamefont{Pfleiderer,
  McMullan, Julian, and Lonzarich}}]{Pfleiderer_et_al_1997}
\bibinfo{author}{\bibfnamefont{C.}~\bibnamefont{Pfleiderer}},
  \bibinfo{author}{\bibfnamefont{G.~J.} \bibnamefont{McMullan}},
  \bibinfo{author}{\bibfnamefont{S.~R.} \bibnamefont{Julian}},
  \bibnamefont{and} \bibinfo{author}{\bibfnamefont{G.~G.}
  \bibnamefont{Lonzarich}}, \bibinfo{journal}{Phys. Rev. B}
  \textbf{\bibinfo{volume}{55}}, \bibinfo{pages}{8330} (\bibinfo{year}{1997}).

\bibitem[{\citenamefont{Ishikawa et~al.}(1976)\citenamefont{Ishikawa, Tajima,
  Bloch, and Roth}}]{Ishikawa_et_al_1976}
\bibinfo{author}{\bibfnamefont{Y.}~\bibnamefont{Ishikawa}},
  \bibinfo{author}{\bibfnamefont{K.}~\bibnamefont{Tajima}},
  \bibinfo{author}{\bibfnamefont{D.}~\bibnamefont{Bloch}}, \bibnamefont{and}
  \bibinfo{author}{\bibfnamefont{M.}~\bibnamefont{Roth}},
  \bibinfo{journal}{Solid State Commun.} \textbf{\bibinfo{volume}{19}},
  \bibinfo{pages}{525} (\bibinfo{year}{1976}).

\bibitem[{\citenamefont{Bak and Jensen}(1980)}]{Bak_Jensen_1980}
\bibinfo{author}{\bibfnamefont{P.}~\bibnamefont{Bak}} \bibnamefont{and}
  \bibinfo{author}{\bibfnamefont{M.~H.} \bibnamefont{Jensen}},
  \bibinfo{journal}{J. Phys. C} \textbf{\bibinfo{volume}{13}},
  \bibinfo{pages}{L881} (\bibinfo{year}{1980}).

\bibitem[{\citenamefont{Nakanishi et~al.}(1980)\citenamefont{Nakanishi, Yanase,
  Hasegawa, and Kataoka}}]{Nakanishi_et_al_1980}
\bibinfo{author}{\bibfnamefont{O.}~\bibnamefont{Nakanishi}},
  \bibinfo{author}{\bibfnamefont{A.}~\bibnamefont{Yanase}},
  \bibinfo{author}{\bibfnamefont{A.}~\bibnamefont{Hasegawa}}, \bibnamefont{and}
  \bibinfo{author}{\bibfnamefont{M.}~\bibnamefont{Kataoka}},
  \bibinfo{journal}{Solid State Commun.} \textbf{\bibinfo{volume}{35}},
  \bibinfo{pages}{995} (\bibinfo{year}{1980}).

\bibitem[{\citenamefont{Dzyaloshinski}(1958)}]{Dzyaloshinski_1958}
\bibinfo{author}{\bibfnamefont{I.~E.} \bibnamefont{Dzyaloshinski}},
  \bibinfo{journal}{J. Phys. Chem. Solids} \textbf{\bibinfo{volume}{4}},
  \bibinfo{pages}{241} (\bibinfo{year}{1958}).

\bibitem[{\citenamefont{Moriya}(1960)}]{Moriya_1960}
\bibinfo{author}{\bibfnamefont{T.}~\bibnamefont{Moriya}},
  \bibinfo{journal}{Phys. Rev.} \textbf{\bibinfo{volume}{120}},
  \bibinfo{pages}{91} (\bibinfo{year}{1960}).

\bibitem[{\citenamefont{Pfleiderer et~al.}(2004)\citenamefont{Pfleiderer,
  Reznik, Pintschovius, v.~L{\"o}hneysen, Garst, and
  Rosch}}]{Pfleiderer_et_al_2004}
\bibinfo{author}{\bibfnamefont{C.}~\bibnamefont{Pfleiderer}},
  \bibinfo{author}{\bibfnamefont{D.}~\bibnamefont{Reznik}},
  \bibinfo{author}{\bibfnamefont{L.}~\bibnamefont{Pintschovius}},
  \bibinfo{author}{\bibfnamefont{H.}~\bibnamefont{v.~L{\"o}hneysen}},
  \bibinfo{author}{\bibfnamefont{M.}~\bibnamefont{Garst}}, \bibnamefont{and}
  \bibinfo{author}{\bibfnamefont{A.}~\bibnamefont{Rosch}},
  \bibinfo{journal}{Nature} \textbf{\bibinfo{volume}{427}},
  \bibinfo{pages}{227} (\bibinfo{year}{2004}).

\bibitem[{\citenamefont{Pfleiderer et~al.}(2001)\citenamefont{Pfleiderer,
  Julian, and Lonzarich}}]{Pfleiderer_Julian_Lonzarich_2001}
\bibinfo{author}{\bibfnamefont{C.}~\bibnamefont{Pfleiderer}},
  \bibinfo{author}{\bibfnamefont{S.~R.} \bibnamefont{Julian}},
  \bibnamefont{and} \bibinfo{author}{\bibfnamefont{G.~G.}
  \bibnamefont{Lonzarich}}, \bibinfo{journal}{Nature}
  \textbf{\bibinfo{volume}{414}}, \bibinfo{pages}{427} (\bibinfo{year}{2001}).

\bibitem[{\citenamefont{Doiron-Leyraud
  et~al.}(2003)\citenamefont{Doiron-Leyraud, Walker, Taillefer, Steiner,
  Julian, and Lonzarich}}]{Doiron-Leyraud_et_al_2003}
\bibinfo{author}{\bibfnamefont{N.}~\bibnamefont{Doiron-Leyraud}},
  \bibinfo{author}{\bibfnamefont{I.}~\bibnamefont{Walker}},
  \bibinfo{author}{\bibfnamefont{L.}~\bibnamefont{Taillefer}},
  \bibinfo{author}{\bibfnamefont{M.~J.} \bibnamefont{Steiner}},
  \bibinfo{author}{\bibfnamefont{S.~R.} \bibnamefont{Julian}},
  \bibnamefont{and} \bibinfo{author}{\bibfnamefont{G.~G.}
  \bibnamefont{Lonzarich}}, \bibinfo{journal}{Nature}
  \textbf{\bibinfo{volume}{425}}, \bibinfo{pages}{595} (\bibinfo{year}{2003}).

\bibitem[{\citenamefont{Lubensky and Stark}(1996)}]{Lubensky_Stark_1996}
\bibinfo{author}{\bibfnamefont{T.~C.} \bibnamefont{Lubensky}} \bibnamefont{and}
  \bibinfo{author}{\bibfnamefont{H.}~\bibnamefont{Stark}},
  \bibinfo{journal}{Phys. Rev. B} \textbf{\bibinfo{volume}{53}},
  \bibinfo{pages}{714} (\bibinfo{year}{1996}).

\bibitem[{\citenamefont{Anisimov et~al.}(1998)\citenamefont{Anisimov, Agayan,
  and Collings}}]{Anisimov_Agayan_Collings_1998}
\bibinfo{author}{\bibfnamefont{M.~A.} \bibnamefont{Anisimov}},
  \bibinfo{author}{\bibfnamefont{V.~A.} \bibnamefont{Agayan}},
  \bibnamefont{and} \bibinfo{author}{\bibfnamefont{P.~J.}
  \bibnamefont{Collings}}, \bibinfo{journal}{Phys. Rev. E}
  \textbf{\bibinfo{volume}{57}}, \bibinfo{pages}{582} (\bibinfo{year}{1998}).

\bibitem[{\citenamefont{Englert et~al.}(2000)\citenamefont{Englert, Stark,
  Longa, and Trebin}}]{Englert_et_al_2000}
\bibinfo{author}{\bibfnamefont{J.}~\bibnamefont{Englert}},
  \bibinfo{author}{\bibfnamefont{H.}~\bibnamefont{Stark}},
  \bibinfo{author}{\bibfnamefont{L.}~\bibnamefont{Longa}}, \bibnamefont{and}
  \bibinfo{author}{\bibfnamefont{H.-R.} \bibnamefont{Trebin}},
  \bibinfo{journal}{Phys. Rev. E} \textbf{\bibinfo{volume}{61}},
  \bibinfo{pages}{2759} (\bibinfo{year}{2000}).

\bibitem[{\citenamefont{Longa et~al.}(2003)\citenamefont{Longa, Ciesla, and
  Trebin}}]{Longa_Ciesla_Trebin_2003}
\bibinfo{author}{\bibfnamefont{L.}~\bibnamefont{Longa}},
  \bibinfo{author}{\bibfnamefont{M.}~\bibnamefont{Ciesla}}, \bibnamefont{and}
  \bibinfo{author}{\bibfnamefont{H.-R.} \bibnamefont{Trebin}},
  \bibinfo{journal}{Phys. Rev. E} \textbf{\bibinfo{volume}{67}},
  \bibinfo{pages}{061705} (\bibinfo{year}{2003}).

\bibitem[{\citenamefont{Ciesla and Longa}(2004)}]{Ciesla_Longa_2004}
\bibinfo{author}{\bibfnamefont{M.}~\bibnamefont{Ciesla}} \bibnamefont{and}
  \bibinfo{author}{\bibfnamefont{L.}~\bibnamefont{Longa}},
  \bibinfo{journal}{Phys. Rev. E} p. \bibinfo{pages}{012701}
  (\bibinfo{year}{2004}).

\bibitem[{\citenamefont{Hertz}(1976)}]{Hertz_1976}
\bibinfo{author}{\bibfnamefont{J.}~\bibnamefont{Hertz}},
  \bibinfo{journal}{Phys. Rev. B} \textbf{\bibinfo{volume}{14}},
  \bibinfo{pages}{1165} (\bibinfo{year}{1976}).

\bibitem[{non()}]{nonanalyticity_footnote}
\bibinfo{note}{We ignore the nonanalytic wave number dependence of $\Gamma$
  that is known to be important for quantum ferromagnets
  \cite{Belitz_et_al_2001a, Belitz_et_al_2001b}; their interplay with the
  chiral DM term has been studied by Vojta and Sknepnek
  \cite{Vojta_Sknepnek_2001}.}

\bibitem[{\citenamefont{Chaikin and Lubensky}(1995)}]{Chaikin_Lubensky_1995}
\bibinfo{author}{\bibfnamefont{P.}~\bibnamefont{Chaikin}} \bibnamefont{and}
  \bibinfo{author}{\bibfnamefont{T.~C.} \bibnamefont{Lubensky}},
  \emph{\bibinfo{title}{Principles of Condensed Matter Physics}}
  (\bibinfo{publisher}{Cambridge University, Cambridge}, \bibinfo{year}{1995}).

\bibitem[{neu()}]{neutron_scattering_footnote}
\bibinfo{note}{Ref.\ \onlinecite{Pfleiderer_et_al_2004} measured a quasi-static
  scattering cross-section that cannot be directly compared to $\chi({\bm q})$.
  Preliminary frequency-integrated results are similar as a function of the
  wave number (C. Pfleiderer, private communication), but more information is
  needed to see whether theory and experiment are at least qualitatively
  compatible.}

\bibitem[{pea()}]{peak_footnote}
\bibinfo{note}{If $\psi$ is increased with everything else held fixed, the peak
  may become more or less sharp, depending on the relative values of $g_1$ and
  $g_2$. One has to keep in mind, however, that $g_2$ increases the effective
  distance $t_{\text R}$ from the ordered phase and counteracts the chiral
  condensation tendency due to $g_1$. A first-order condensation transition can
  therefore be expected to take place only if the fully renormalized $g_2$ is
  small compared to $g_1$ in natural units.}

\bibitem[{\citenamefont{Wright and Mermin}(1989)}]{Wright_Mermin_1989}
\bibinfo{author}{\bibfnamefont{D.~C.} \bibnamefont{Wright}} \bibnamefont{and}
  \bibinfo{author}{\bibfnamefont{N.~D.} \bibnamefont{Mermin}},
  \bibinfo{journal}{Rev. Mod. Phys.} \textbf{\bibinfo{volume}{61}},
  \bibinfo{pages}{385} (\bibinfo{year}{1989}).

\bibitem[{\citenamefont{Crooker}(2001)}]{Crooker_2001}
\bibinfo{author}{\bibfnamefont{P.~P.} \bibnamefont{Crooker}}, in
  \emph{\bibinfo{booktitle}{Chirality in Liquid Crystals}}, edited by
  \bibinfo{editor}{\bibfnamefont{H.~S.} \bibnamefont{Kitzerow}}
  \bibnamefont{and} \bibinfo{editor}{\bibfnamefont{C.}~\bibnamefont{Bahr}}
  (\bibinfo{publisher}{Springer, New York}, \bibinfo{year}{2001}), p.
  \bibinfo{pages}{186}.

\bibitem[{\citenamefont{Belitz et~al.}(2001{\natexlab{a}})\citenamefont{Belitz,
  Kirkpatrick, Mercaldo, and Sessions}}]{Belitz_et_al_2001a}
\bibinfo{author}{\bibfnamefont{D.}~\bibnamefont{Belitz}},
  \bibinfo{author}{\bibfnamefont{T.~R.} \bibnamefont{Kirkpatrick}},
  \bibinfo{author}{\bibfnamefont{M.~T.} \bibnamefont{Mercaldo}},
  \bibnamefont{and} \bibinfo{author}{\bibfnamefont{S.~L.}
  \bibnamefont{Sessions}}, \bibinfo{journal}{Phys. Rev. B}
  \textbf{\bibinfo{volume}{63}}, \bibinfo{pages}{174427}
  (\bibinfo{year}{2001}{\natexlab{a}}).

\bibitem[{\citenamefont{Belitz et~al.}(2001{\natexlab{b}})\citenamefont{Belitz,
  Kirkpatrick, Mercaldo, and Sessions}}]{Belitz_et_al_2001b}
\bibinfo{author}{\bibfnamefont{D.}~\bibnamefont{Belitz}},
  \bibinfo{author}{\bibfnamefont{T.~R.} \bibnamefont{Kirkpatrick}},
  \bibinfo{author}{\bibfnamefont{M.~T.} \bibnamefont{Mercaldo}},
  \bibnamefont{and} \bibinfo{author}{\bibfnamefont{S.~L.}
  \bibnamefont{Sessions}}, \bibinfo{journal}{Phys. Rev. B}
  \textbf{\bibinfo{volume}{63}}, \bibinfo{pages}{174428}
  (\bibinfo{year}{2001}{\natexlab{b}}).

\bibitem[{\citenamefont{Vojta and Sknepnek}(2001)}]{Vojta_Sknepnek_2001}
\bibinfo{author}{\bibfnamefont{T.}~\bibnamefont{Vojta}} \bibnamefont{and}
  \bibinfo{author}{\bibfnamefont{R.}~\bibnamefont{Sknepnek}},
  \bibinfo{journal}{Phys. Rev. B} \textbf{\bibinfo{volume}{64}},
  \bibinfo{pages}{052404} (\bibinfo{year}{2001}).

\end{thebibliography}

\end{document}